\documentclass[11pt,twoside]{article}

\usepackage{asp2006}
\usepackage{epsf}
\usepackage{lscape}

\begin{document}

\title{O(He) Stars}
\author{T\@. Rauch$^1$, E\@. Reiff$^1$, K\@. Werner$^1$, J\@. W\@. Kruk$^2$}
\affil{$^1$Institute for Astronomy and Astrophysics, 
           Kepler Center for Astro and Particle Physics, 
           Eberhard Karls University, T\"ubingen, Germany}    
\affil{$^2$Department of Physics and Astronomy, Johns Hopkins University, Baltimore, U.S.A.}    

\begin{abstract}
Spectral analyses of H-deficient post-AGB stars have shown that a small 
group of four extremely hot objects exists which have almost pure He 
absorption-line spectra in the optical. These are classified as O(He) stars. 
For their evolution there are two scenarios: They could be the long-sought 
hot successors of RCrB stars, which have not been identified up to now. If this 
turns out to be true, then a third post-AGB evolutionary sequence is revealed, which 
is probably the result of a double-degenerate merging process.
An alternative explanation might be that O(He) stars are post early-AGB stars.
These depart from the AGB just before they experience their first thermal pulse (TP)
which will then occur as a late thermal pulse (LTP). This would be a link
to the low-mass He-enriched sdO stars and low-mass, particularly He-rich PG\,1159 stars. 
\end{abstract}

\section{Introduction}
\label{sect:introduction}
95\% of all stars end their lives as white dwarfs. About 20\% of the
hot post-AGB stars are H deficient. Most of these are the result of a
(very) late He-shell flash, but the evolutionary status of a fraction of about
$10-20$\% of the hottest H-deficient stars, namely the O(He) stars
\citep{m1991}, is as yet unexplained. O(He) stars form a small group of four extremely 
hot post-AGB stars which exhibit absorption-line spectra dominated by He\,{\sc ii}. For a 
detailed introduction see \citet{rdw1998} and \citet{rea2006}.

(V)LTP evolutionary models can explain the observed He/C/O abundances in 
Wolf-Rayet and PG1159 stars \citep[see, e.g.,][]{hea1999}, but they have never 
reproduced He-dominated surface abundances. Recently, Miller Bertolami \& Althaus (2006) presented 
a ``numerical test'' where they increased the mass-loss rate of a LTP star 
artificially in order to ``blow away'' the H from the stellar surface. However, 
this is in contradiction with radiation-driven wind theory \citep{pea1988} and, 
if at all, valid for low-mass O(He) stars only (see below).

As an alternative, a third post-AGB evolutionary sequence might exist and the 
O(He) stars could be the successors of the RCrB stars which are relatively cool 
($T_\mathrm{eff}$ around 10\,000\,K) stars with He-dominated atmospheres, too. 
If this is true, we can expect similar metal abundances. In order to investigate 
on these in O(He) stars, we have performed FUV observations of all four objects with 
the Far Ultraviolet Spectroscopic Explorer (FUSE).

\section{Observations and Analysis}
\label{sect:obsana}
Previous spectral analyses of O(He) stars were based on optical, UV (International
Ultraviolet Explorer, IUE), and X-ray (ROSAT) observations \citep[][1996, 1998]{rkw1994}.
The photospheric parameters are summarized in Table~\ref{tab:php}.

\begin{table}[ht]
\caption[]{Parameters of the four known O(He) stars, determined by our analyses
of optical spectra \citep{rdw1998}. Typical uncertainties are: 
$T_\mathrm{eff}$ $\raisebox{0.20em}{{\tiny \hspace{0.2mm}\mbox{$\pm$}\hspace{0.2mm}}}$10\,\%, $\log g$
$\raisebox{0.20em}{{\tiny \hspace{0.2mm}\mbox{$\pm$}\hspace{0.2mm}}}$0.5\,dex, abundance ratios 
$\raisebox{0.20em}{{\tiny \hspace{0.2mm}\mbox{$\pm$}\hspace{0.2mm}}}$0.3\,dex. For comparison, the last two
lines give the mean element abundances of the majority RCrB stars and the
peculiar RCrB star V854~Cen, respectively \citep{rl1996}. The scatter
around the mean C, N, O, and Si abundances is 
$\raisebox{0.20em}{{\tiny \hspace{0.2mm}\mbox{$\pm$}\hspace{0.2mm}}}$0.15,
$\raisebox{0.20em}{{\tiny \hspace{0.2mm}\mbox{$\pm$}\hspace{0.2mm}}}$0.21, 
$\raisebox{0.20em}{{\tiny \hspace{0.2mm}\mbox{$\pm$}\hspace{0.2mm}}}$0.46,
$\raisebox{0.20em}{{\tiny \hspace{0.2mm}\mbox{$\pm$}\hspace{0.2mm}}}$0.18\,dex, respectively.}\vspace{-2mm}
\label{tab:php}
\begin{center}
\begin{tabular}{lcr@{.}lr@{.}lr@{.}lr@{.}lr@{.}l}
\hline\noalign{\smallskip}
\multicolumn{1}{c}{} &
\multicolumn{1}{c}{$T_\mathrm{eff}$} &
\multicolumn{2}{c}{$\log g$} &
\multicolumn{2}{c}{H\,/\,He} &
\multicolumn{2}{c}{C\,/\,He} &
\multicolumn{2}{c}{N\,/\,He} &
\multicolumn{2}{c}{O\,/\,He}  \\
\cline{5-12}
\noalign{\smallskip}
\multicolumn{1}{c}{} &
\multicolumn{1}{c}{kK} &
\multicolumn{2}{c}{cgs} &
\multicolumn{8}{c}{number ratio} \\
\noalign{\smallskip}
\hline
\noalign{\smallskip}
LoTr\,4  & 120 &  5&5 &      0&5 & \raisebox{0.20em}{{\tiny \mbox{$<$}}} 0&004 & 0&001 & \raisebox{0.20em}{{\tiny \mbox{$<$}}} 0&008       \\
HS\,1522+6615 & 140 &  5&5 &      0&1 &      0&003 & \multicolumn{4}{c}{}           \\
HS\,2209+8229 & 100 &  6&0 & \hspace{2mm}\raisebox{0.20em}{{\tiny \mbox{$<$}}} 0&2 & \multicolumn{6}{c}{} \\
K\,1-27  & 105 & ~6&5 & \hbox{}\hspace{2mm}\raisebox{0.20em}{{\tiny \mbox{$<$}}} 0&2 & \raisebox{0.20em}{{\tiny \mbox{$<$}}} 0&005 & 0&005 & \multicolumn{2}{c}{} \\
\noalign{\smallskip}
\hline
\noalign{\smallskip}
majority RCrB &          &   \multicolumn{2}{c}{}  & \hspace{2mm}\raisebox{0.20em}{{\tiny \mbox{$<$}}} 0&0001&    0&010 & 0&004 & 0&005 \\
V854 Cen      &          &   \multicolumn{2}{c}{}  &      0&5   &    0&030 & 0&0003& 0&003\\
\noalign{\smallskip}
\hline
\end{tabular}
\end{center}
\end{table}

Since high-resolution and high-S/N UV spectra are a prerequisite for
a reliable determination of metal abundances in hot stars, we have applied for
observations with the Hubble Space Telescope (HST) and the
Space Telescope Imaging Spectrograph (STIS). We were awarded observational time
in Cycle 13. Unfortunately, STIS encountered a failure during that Cycle on 
Aug 4, 2004 while our first observations were scheduled on Aug 9. 

A FUSE campaign in 2002 (Cycle 3, program C178, exposure time about 40\,ksec) 
was more successful. For the analysis of the obtained FUV spectra we used 
{\sc TMAP}, the T\"ubingen NLTE Model-Atmosphere Package \citep{wea2003}, and 
calculated fully-line blanketed models in order to identify weak photospheric 
lines. In  the course of this analysis, it turned out that the spectra
are strongly contaminated by interstellar absorption and the S/N was not 
sufficient for our purpose. The mass-loss rates could be determined from the 
O\,{\sc vi} $\lambda\lambda$\,1031.9, 1037.6\,\AA\ resonance doublet \citep{rea2006} 
and appear in agreement with radiation-driven wind theory. 

One O(He) star (HS\,1522+6615) was included in a re-observation campaign
(Cycle 7, program U103, exposure time about 4\,ksec) and our own re-observation 
proposal (Cycle 8, program H024, 204\,ksec in order to improve the S/N) was accepted.
These observations were scheduled for summer 2007 -- the FUSE failure on July 12, 2007 
prevented them. Thus, at the moment, we cannot determine the photospheric metal abundances 
of the O(He) stars reliably.

\section{Evolutionary Scenarios and Conclusions}
\label{sect:conclusions}
The group of four known O(He) stars divides into two sub-groups,
    two high-mass 
($M\approx 0.8\,\mathrm{M_\odot}$, the central star (CS) of the planetary nebula (PN) LoTr\,4 and HS\,1522+6615) 
and two  low-mass 
($M\approx 0.5\,\mathrm{M_\odot}$, CSPN of K\,1$-$27 and HS\,2209+8829) members.
Its is possible that different evolutionary scenarios are valid for each group.
 
\begin{figure}[ht]
\epsfxsize=\textwidth
\epsffile{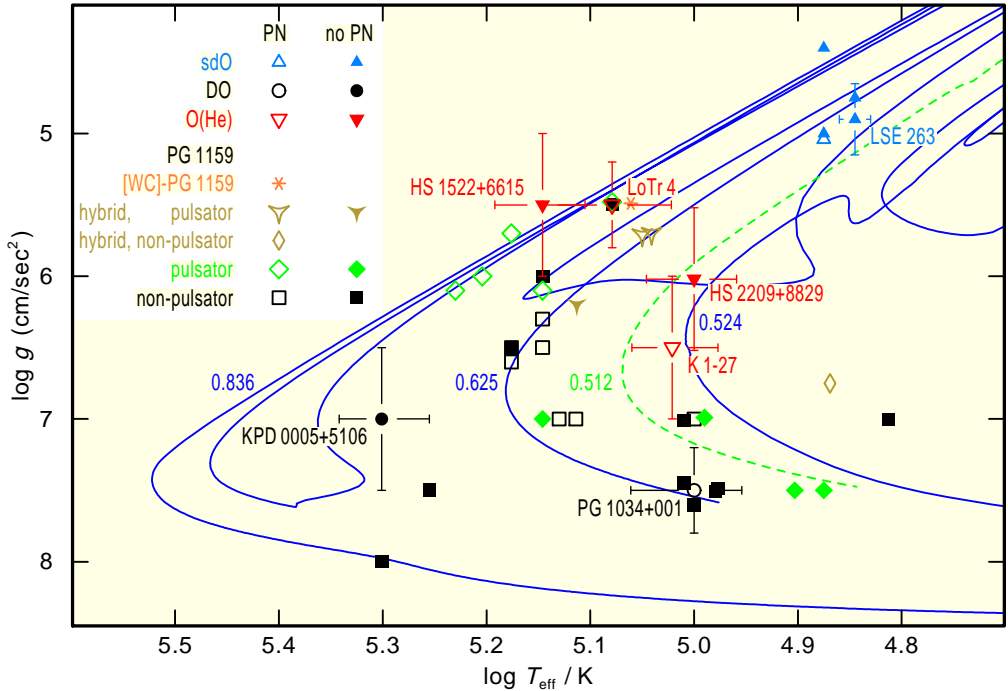}
\caption{
Positions of the four known O(He) stars and related objects in
the $\log T_\mathrm{eff}$\,--\,$\log g$ diagram compared to theoretical evolutionary
tracks for born-again post-ABG stars of \citet[][full lines]{b1995} and
\citet[][dashed]{ma2006}. The tracks are labeled
with the respective stellar masses (in M$_\odot$).
}
\label{fig:teffg}
\end{figure}

The low-mass O(He) stars may be so-called post early-AGB stars, i.e\@. they
experienced the first thermal pulse (TP) after their departure from the asymptotic giant branch (AGB). 
\citet{ma2006} calculated an evolutionary track for a $M = 0.512\,\mathrm{M_\odot}$ 
post early-AGB star which closely matches the positions 
of the CSPN of K\,1$-$27 and of HS\,2209+822 in the $\log T_\mathrm{eff}-\log g$ 
plane (Fig.~\ref{fig:teffg}). 
Due to an artificially increased mass-loss rate (ten times higher than predicted by 
radiation-driven wind theory), a strong H deficiency is achieved. Spectral analysis 
of the FUSE spectra has shown that the mass-loss rates of O(He) stars are not higher 
than predicted by radiation-driven wind theory \citep{pea1988} and therefore, any 
change of the surface composition due to a stellar wind is unlikely. Moreover, for 
the H abundance in these O(He) stars, presently only upper limits are known 
(Tab.~\ref{tab:php}). Therefore a further, detailed investigation on a distinct 
``sdO(He) $\rightarrow$ O(He) $\rightarrow$ DO white dwarf'' evolutionary channel
\citep[cf\@.][]{rdw1998} is necessary. However, it is worth to note that the
DO-type CSPN PG\,1034+001 \citep[cf.][Fig.~\ref{fig:teffg}]{wdw1995} may be a descendant 
from low-mass O(He) or PG\,1159 stars.

The high-mass O(He) stars appear to follow the ``normal'' born-again scenario \citep{iea1983}
where a (very) late TP causes the H-deficiency. However, it is worthwhile to note that
both known objects, the CSPN of LoTr\,4 as well as HS\,1522+6615, show remaining H in 
their spectra (Tab.~\ref{tab:php}). Since their surface gravity $g$ is relatively low, 
gravitational settling during their following evolution might turn them into H-rich
(DA) white dwarfs. Therefore, it is unclear whether the extremely hot DO white dwarf 
KPD\,0005+5106 (Fig.~\ref{fig:teffg}, cf\@. Werner et al\@. these proceedings) is
a successor of high-mass O(He) stars.

An alternative O(He) scenario is that of double-degenerate mergers. Similar
H/He surface compositions suggest that O(He) stars are the progeny of RCrB
stars \citep{rea2006} and they follow a
``RCrB $\rightarrow$ O(He) $\rightarrow$ DO white dwarf'' sequence.  
Available FUSE spectra of O(He) stars do not show isolated metal 
lines and thus, allow to give upper limits for metal abundances only. 
Iron-group abundances are probably solar. 

In order to make progress, further UV observations with COS or STIS are highly 
desirable in order to determine C, N, O, and Si abundances precisely in order 
to corroborate a possible link to RCrBs.
\vspace{5mm}

\acknowledgements
T.R\@. is supported by the \emph{German Astrophysical Virtual Observatory} project
of the German Federal Ministry of Education and Research (BMBF) under grant 05\,AC6VTB. 
E.R\@. is supported by DFG grant We1312/30$-$1. 
J.W.K\@. is supported by the FUSE project, funded by NASA contract NAS5$-$32985.

\end{document}